\documentclass[12pt]{spieman}  
\usepackage[utf8]{inputenc}
\usepackage{amsmath,amsfonts,amssymb}
\usepackage{graphicx}
\usepackage{hyperref}
\usepackage{setspace}
\usepackage{tocloft}
\usepackage{float}
\usepackage[left]{lineno}
\title{PIRATES - a machine-learning framework for polarized, interferometric image reconstruction}

\author[a,b]{Lucinda Lilley *}
\author[a,b,c]{Barnaby Norris}
\author[a,b]{Peter Tuthill}
\author[a]{Eckhart Spalding}
\author[d,e]{Miles Lucas}
\author[i]{Manxuan Zhang}
\author[i]{Maxwell Millar-Blanchaer}
\author[m]{Christophe Pinte}
\author[d]{Michael Bottom}
\author[e, f,g,h]{Olivier Guyon}
\author[e]{Julien Lozi}
\author[e,k]{Vincent Deo}
\author[e,h,n]{S\'ebastien Vievard}
\author[l,a,b]{Alison P Wong}
\author[j, e]{Kyohoon Ahn}
\author[f]{Jaren Ashcraft}

\vspace{10pt}

\affil[a]{{\small Sydney Institute for Astronomy, School of Physics, 
Physics Road, University of Sydney, NSW 2006, Australia}}
\affil[b]{{\small Sydney Astrophotonic Instrumentation Laboratories, Physics Road, University of Sydney, NSW 2006, Australia}}
\affil[c]{{\small AAO-USyd, School of Physics, University of Sydney, NSW 2006, Australia}}
\affil[d]{{\small Institute for Astronomy, University of Hawai'i, 640 N. Aohoku Pl., Hilo, HI 96720, USA}}
\affil[e]{{\small Subaru Telescope, National Astronomical Observatory of Japan, 650 N. Aohoku Pl., Hilo, HI 96720, USA}}
\affil[f]{{\small College of Optical Sciences, University of Arizona, Tucson, AZ 87521, USA}}
\affil[g]{{\small Steward Observatory, University of Arizona, Tucson, AZ 87521, USA}}
\affil[h]{{\small Astrobiology Center, 2 Chome-21-1, Osawa, Mitaka, Tokyo, 181-8588, Japan}}
\affil[i]{{\small Department of Physics, University of California, Santa Barbara, CA, 93106, USA}}
\affil[j]{{\small Technology Center for Astronomy and Space Science, Korea Astronomy and Space Science Institute, 776 Daedeok-daero, Yuseong-gu, Daejeon 34055, Republic of Korea}}
\affil[k]{{\small Optical Sharpeners SAS, France}}
\affil[l]{{\small Discipline of Business Analytics, University of Sydney Business School, University of Sydney, NSW 2006, Australia}}
\affil[m]{{\small Univ. Grenoble Alpes, CNRS, IPAG, 38000 Grenoble, France}}
\affil[n]{{\small Space Science and Engineering Initiative, College of Engineering, Institute for Astronomy, University of Hawaii, 640 North Aohoku Place, Hilo, HI, 96720, USA}}

\cftpagenumbersoff{figure}
\cftpagenumbersoff{table} 
\begin{document} 
\maketitle

\begin{abstract}
 
Optical interferometric image reconstruction is a challenging, ill-posed optimization problem which usually relies on heavy regularization for convergence. Conventional algorithms regularize in the pixel domain, without cognizance of spatial relationships or physical realism, with limited utility when this information is needed to reconstruct images. Here we present PIRATES (Polarimetric Image Reconstruction AI for Tracing Evolved Structures), the first image reconstruction algorithm for optical polarimetric interferometry. PIRATES has a dual structure optimized for parsimonious reconstruction of high fidelity polarized images and accurate reproduction of interferometric observables. The first stage, a convolutional neural network (CNN), learns a physically meaningful prior of self-consistent polarized scattering relationships from radiative transfer images. The second stage, an iterative fitting mechanism, uses the CNN as a prior for subsequent refinement of the images with respect to their polarized interferometric observables. Unlike the pixel-wise adjustments of traditional image reconstruction codes, PIRATES reconstructs images in a latent feature space, imparting a structurally derived implicit regularization. We demonstrate that PIRATES can reconstruct high fidelity polarized images of a broad range of complex circumstellar environments, in a physically meaningful and internally consistent manner, and that latent space regularization can effectively regularize reconstructed images in the presence of realistic noise.

\end{abstract}


{\noindent \footnotesize\textbf{*} Correspondence to: Lucinda Lilley,  \linkable{lucinda.lilley@sydney.edu.au} }


\section{Introduction}
\label{sect:intro}  

Image reconstruction is an essential technique for deriving physical meaning from interferometric observables and is complementary to forwards modeling that yields plausible descriptions of an astrophysical scene. Reconstructing an accurate image from interferometric observables (visibilities and closure phases) is an inherently ill-posed and non-convex optimization problem \cite{Thiebaut2009}. The reconstruction process requires the information contained in interferometric observables to be re-represented in an over-sampled basis (the image), with a significantly higher number of degrees of freedom (pixels) than there are observables - a task with a vast number of degenerate solutions, most of which will be un-physical or dominated by spurious structure or noise. To increase the probability of obtaining physically meaningful solutions, the convergence of an image reconstruction algorithm is often heavily driven by regularization penalties, which place constraints on the statistics of the reconstructed images.  Commonly used regularization penalties include; maximum entropy, total variation, L1 and L2 norms \cite{Buscher1994, Thiebaut2008, Baron2010}. These forms of pixel-distribution regularization typically aggregate some function of the pixel values, losing information on spatial connectivity and two dimensional structure. In addition to regularization, a prior on the flux distribution can be specified, for example that flux is expected in the image center \cite{Thiebaut2017}. With the assistance of regularization and a prior, image reconstruction algorithms aim to converge on a reconstructed image that is meaningful, contains minimal spurious noise, and most importantly, does not mislead the viewer to infer more complex phenomena than may be mathematically constrained by the original observables.

\vspace{3mm}

There are many successful image reconstruction algorithms for un-polarized, optical interferometry \cite{Buscher1994, Thiebaut2008, Baron2010, Thiebaut2010, Thiebaut2017, Baron2016}, however no algorithms currently exist for optical, polarimetric interferometry. The polarization state of light contains critical information otherwise not captured by standard, unpolarized interferometry, and major facilities are now looking towards fully leveraging this information to broaden the observational reach of new and existing instruments \cite{Setterholm2020, slayman}. Polarized interferometric image reconstruction has an important additional challenge compared to unpolarized interferometric image reconstruction: the images of the different polarization components (e.g. Stokes I, Q and U) are more accurately described as spatial maps of vector components, which at optical wavelengths define the geometry of light-scattering processes. As such, the Stokes images require simultaneous reconstruction, and must mutually depict a physically realistic and internally consistent description of the physics of the scattered scene. Regularizing for these physical constraints from the distribution of flux within and between each Stokes image is a challenging task without a straightforward solution, as each pixel has a complex relationship with its surrounding pixels, both spatially and across each polarization vector component. Polarimetric images of Stokes Q and U also break the assumption of image positivity which is required for the use of many pixel-distribution regularization penalties like maximum entropy or the L1 and L2 norms\cite{Valenzuela2010, Mus2024}. As such, polarimetric image reconstruction has the simultaneous requirement of a nuanced adaptation of common pixel-distribution regularization penalties, and a way to ensure that reconstructed Stokes image triplets are internally consistent and physically meaningful. 
 
\subsection{The PIRATES algorithm}
Here we present PIRATES, a two stage machine learning framework (pre-trained CNN + iterative fitting), that can handle these complex regularization requirements, parsimoniously reconstructing high fidelity polarized images that accurately match polarized observables (Figure \ref{vid:ml_framework}). 


\begin{video}[htbp]
\centering 
{\includegraphics[trim = 7.8cm 4.2cm 4.45cm 6.85cm, clip,  width=\textwidth]{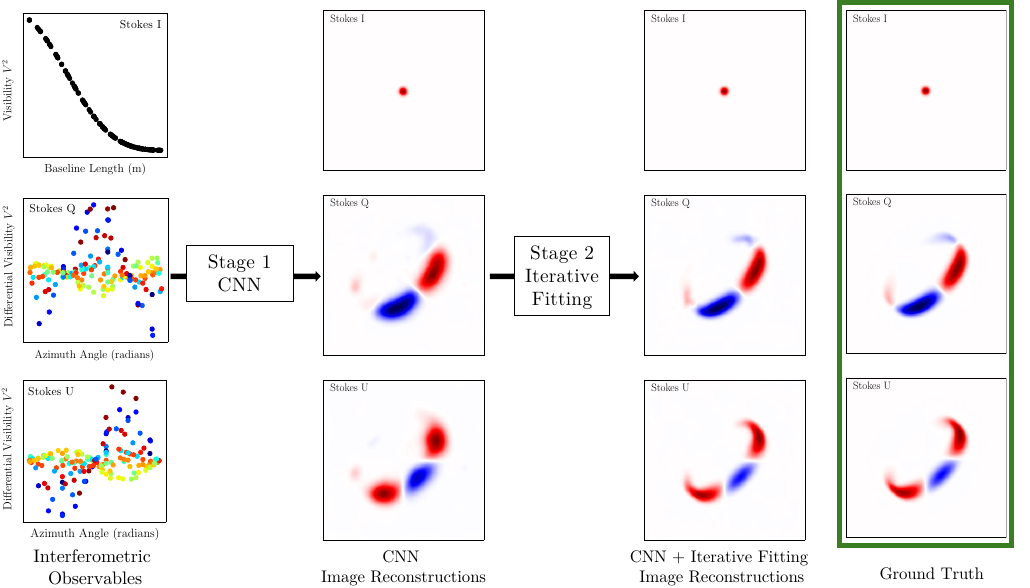}}
\caption{\label{vid:ml_framework} Schematic of PIRATES' dual structure. Stage 1 is a pre-trained CNN which learns the mapping from polarized interferometric observables to the Stokes image triplet (I, Q, U). This mapping between observables and images is learned from `probe' radiative transfer images which are agnostic to specific astrophysical structure. The resulting algorithm then generalizes well to astrophysically meaningful examples like the disk above. Stage 2 is an iterative fitting algorithm - it optimizes the pre-trained CNN images now with respect to their interferometric observables. Changes to the images are made via updates to convolutional kernels in a latent feature space, and not the pixel wise adjustments of conventional unpolarized image reconstruction codes, imparting a structurally derived implicit regularization on reconstructed images.}
\end{video}

Firstly, the radiative transfer code MCFOST \cite{Pinte2006, Pinte2009} is used to probe a space of physically realistic and relevant polarized images of smooth and diffuse circumstellar dust. A distribution of Gaussian probe `volumes' of dust are randomly generated and injected into the model domain, spanning the physical field of view and spatial scales represented by the observables whilst remaining agnostic to specific astrophysical structures. Similarly, a distribution of dust densities, grain sizes, etc. is used to probe the space of valid scattering solutions. Pre-training the CNN on this data teaches it to learn the mapping between polarized images and polarized interferometric observables (visibilities and closure phases), over a space of self-consistent images which have physically meaningful spatio-polarimetric relationships. Rather than using an explicit term, regularization is encoded in the CNN training data - at inference time, reconstructions are implicitly regularized by the statistics of the training distribution. The architecture of a CNN allows the mapping between observables and images to be learned via non-linear operations of convolutional spatial kernels, and not the pixel-level adjustments performed in conventional image reconstruction codes. Whilst the CNN learns from considerable training data, the enforced generalization of convolutional kernels to the whole of each image gives CNNs a powerful, structurally derived inductive bias \cite{Wang2023}. The combination of appropriately regularized pre-training data and the architectural inductive bias of a well built and tuned CNN imparts a powerful implicit regularization on image reconstructions \cite{Ulyanov2017}, in this case towards physically consistent Stokes triplets depicting scattered light.

\vspace{3mm}

At inference time, we use our pre-trained CNN to reconstruct polarimetric images from observed visibilities and closure phases, for a single test example (stage 1). The CNN-predicted image provides a moderate detail and low noise estimate of the reconstructed images with a single pass, and initializes the iterative fitting in a physically meaningful feature space close to the true minimum (Figure \ref{vid:ml_framework}, middle column). Iterative fitting (stage 2) then performs additional refinement to the CNN-predicted images by now optimizing them (in the learned feature-space) with respect to the interferometric observables, making physically meaningful updates to the polarized images without introducing spurious structure or noise into the images (Figure \ref{vid:ml_framework}, right column).

 \vspace{3mm}
\subsection{Differential interferometric polarimetry}
Here we apply PIRATES to simulated polarized-interferometric observables produced by the VAMPIRES Instrument (Visible Aperture Masking Polarimeter for Resolving Exoplanetary Signatures / Evolved Stars), a differential imaging polarimeter mounted on the SCExAO bench of the SUBARU Telescope \cite{Norris2015, slayman}. VAMPIRES is equipped to perform single telescope interferometry via non-redundant aperture masking (NRM). VAMPIRES + NRM is a prototypical example of polarized interferometry - an instrument built on earlier successes from the now decommissioned NACO + SAMPOL/VLT \cite{Lenzen2003, Tuthill2010}, and SUSI \cite{Davis1999}. Operating at optical wavelengths (600 - 800 nm) where polarization signatures induced by dust-scattered starlight are strong \cite{Snik}, VAMPIRES + NRM performs simultaneous multi-wavelength polarimetric interferometry of linear polarization states Stokes I, Q and U, providing a sub-diffraction limited spatio-spectral description of observationally challenging inner-circumstellar environments that lay host to dusty material \cite{Norris2015, Norris2015_thesis}, such as proto-planetary disks and evolved-star mass-loss shells. Within the present work, we demonstrate our algorithm by simulating observations with the 18 hole `g18' aperture mask, which has 153 baseline vectors \cite{Norris2015}. For consistency, the IAU polarimetric coordinate system is adopted throughout this paper \cite{Hamaker1996}.

\vspace{3mm}

The reduced data products for polarimetric interferometry are squared visibilities (or visibility ratios) and closure phases for Stokes $I$, $Q$ and $U$. The Stokes formalism represents each point in a wavefront with a Stokes vector, $\vec{S} = (I, Q, U, V)$, where $I$ is intensity, and $Q$ and $U$ are the orthogonal linear polarization components \cite{Hecht2002}. Circular polarization, $V$ is omitted from the present work (set as $V=0$), as it is not measured by VAMPIRES. For Stokes $I$, visibilities take the conventional format (Figure \ref{vid:vamp_vis}, left panel), however for Stokes $Q$ and $U$, the visibility and closure phase measurements are polarized and differential (Figure \ref{vid:vamp_vis}, middle and right panels). We define the differential visibility as the ratio of orthogonally-polarized raw visibilities, and the differential closure phase as the difference between orthogonally polarized raw closure phases. For VAMPIRES, measurement of Stokes Q and U is the result of a triple-differential measurement - linear polarization states H and V (Stokes Q), and H45 and V45 (Stokes U), are made thrice by modulating three optical components - a ferroelectric liquid crystal (FLC), a half wave plate (HWP) and a polarizing beam-splitter (BS) \cite{slayman}. Triple division of the corresponding visibilities yield very high precision and self-calibrated measurement of Stokes Q and U, and remove the conventional need for a PSF calibrator star \cite{Norris2015}. For description of the VAMPIRES instrument and data reduction process, please see \cite{Norris2015} and \cite{slayman}.

\begin{video}[htbp]
\centering
{\includegraphics[trim = 0.15cm 0cm 0cm 0cm, clip,  width= 0.999\textwidth]{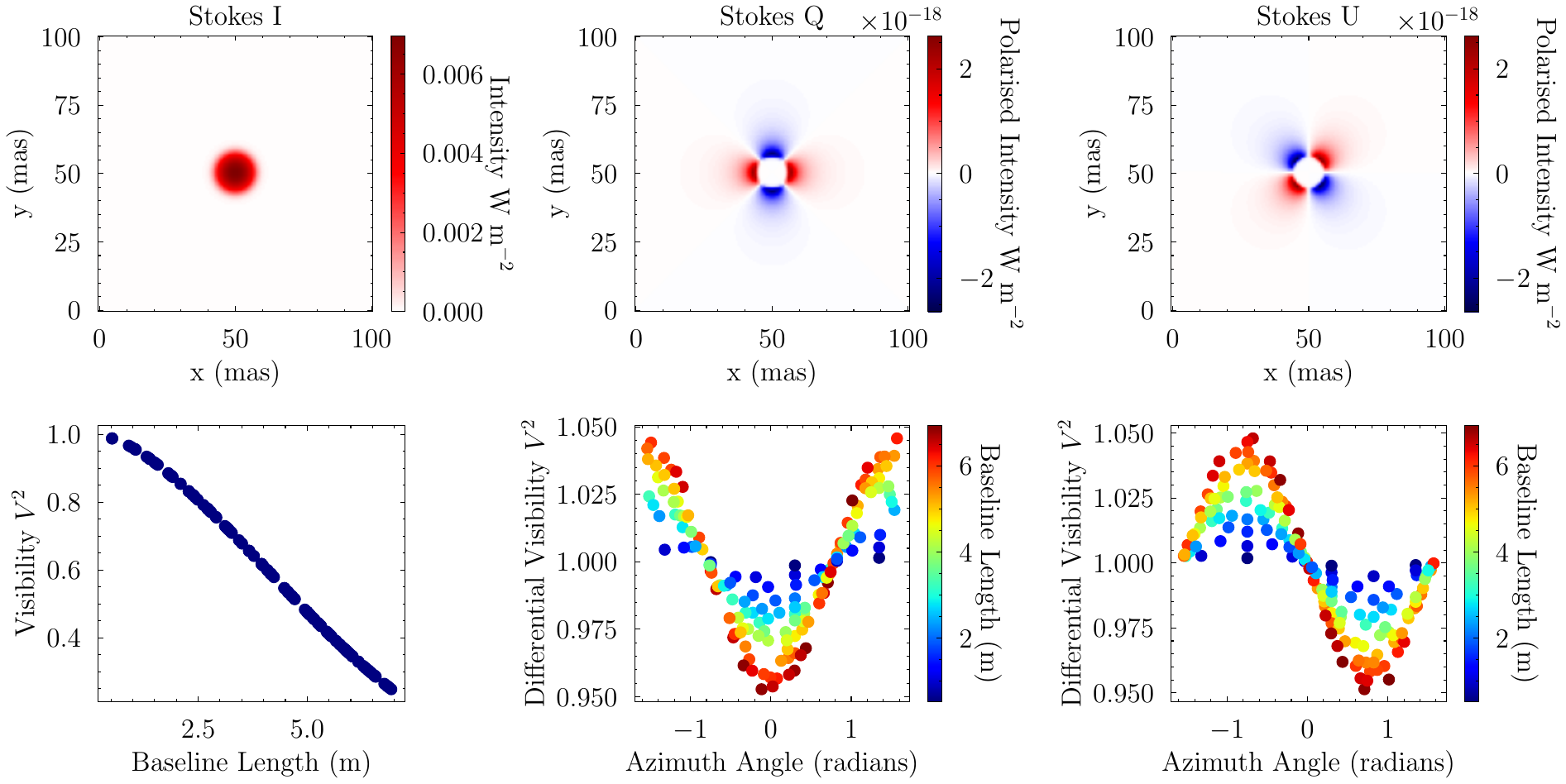}}
\caption{\label{vid:vamp_vis} Simulated VAMPIRES observables for a thick, spherically symmetric circumstellar envelope: images and visibilities in Stokes I, and polarimetric and self-differential images and visibilities in Stokes Q and U. Stokes I visibilities have the traditional interferometric format, whilst Stokes Q and U visibilities are polarized and self-differential. Polarimetric visibilities of symmetric objects have a characteristic sinusoidal and co-sinusoidal appearance for Stokes Q and Stokes U.}
\end{video}

Physical interpretation of polarimetric interferometry data first involves modeling the circumstellar geometry. Historically, this has been achieved via parametric forwards modeling \cite{Ireland2005, Norris2012, Haubois2019}. For computational tractability, several significant physical assumptions are typically made - firstly, that the scattering environment is optically thin and homogeneous \cite{Ireland2005, Norris2012, Haubois2019}. Secondly, models for scattering allow for additional dust grains to increase polarization amplitudes indefinitely: an approximation that eventually becomes unphysical with the onset of effects from multiple scattering events \cite{Doicu2020, Ireland2005, Norris2012, Haubois2019}. The limitations of these assumptions are elucidated by recent VAMPIRES datasets for complex evolved stars \cite{Lilley2025}, for which there are visible and systematic departures from these assumptions. To properly model complex targets, a physically complete treatment of optical depth and inhomogeneity in the dust is required, necessitating full radiative transfer models. Parametric forwards modeling with radiative transfer code is computationally expensive - in contrast to simple parametric scattering models which may run with GPU acceleration in several milli-seconds, a full radiative transfer model may take several minutes. In addition, the design and choice of candidate parametric models becomes counter intuitive when inhomogeneity, optical depth and asymmetry all interact. These limitations and challenges motivate the development of our algorithm, which we design as a tool to reconstruct polarimetric images of challenging and complex circumstellar scenes. PIRATES aims to produce maximally parsimonious and physically self-consistent polarized images which best fit the polarized interferometric observables. The resulting images can also be used to initialize and constrain further parametric model-fitting.

\section{Materials and Methods}\label{sect:materials_methods}  

\vspace{3mm}

\subsection{Algorithm Architecture}

Here we summarize the structure of PIRATES (Figure \ref{vid:ml_framework}), the chosen architecture and hyper-parameters; the tuning performed to obtain these values is explained in Section \ref{sect:results_1_general}. All code within this study is written in Python, neural networks are built in TensorFlow \cite{Tensorflow} and Keras \cite{Keras} and run on an NVIDIA GeForce RTX 4090. 

\begin{video}[h!]
\begin{center} 
{\includegraphics[trim = 1.1cm 6.2cm 5.0cm 5.5cm, clip,  width= 0.99\textwidth]{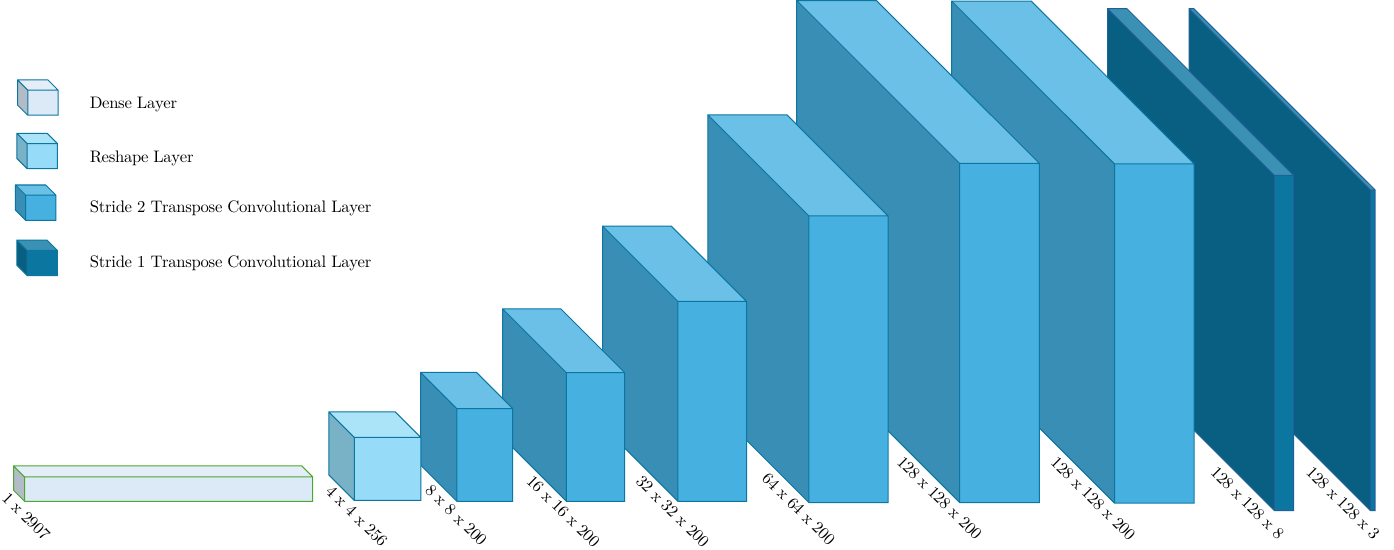}}
\end{center}
\caption{\label{vid:final_F} Structure of PIRATES CNN (stage 1). Inputs to the network are vectors of polarized interferometric observables (visibilities and closure phases) (1, 2907). These are accepted by a dense layer, and then reshaped into a two dimensional image with many channels (4 x 4 x 256), which allows for the subsequent application of transpose convolutional layers. Five stride-two transpose convolutional layers then up-sample until the spatial output size (128 x 128) is achieved, with 200 kernels in each layer. Then follow three stride 1 convolutional layers - the first with 200 kernels, and the last two with 8 and 3 to downsize the output channel dimension to 3 (one for each of I, Q and U). The outputs of the network are the polarized image triplets - Stokes I, Q and U (128 x 128 x 3). Choice of this architecture is discussed in Section \ref{sect:results_1_CNN}.}
\end{video}

Stage 1 of PIRATES is a CNN which takes polarimetric observables as input, and outputs reconstructed polarimetric images (Figure \ref{vid:final_F}). Each training example input consists of a single vector of interferometric observables: visibilities and closure phases for Stokes I, and polarized differential visibilities and closure phases for Stokes Q and U. This vector is then reshaped to have two spatial and one filter dimension, after which it is passed to eight transpose convolutional layers. The first five of these layers have a stride of 2 and up-sample the input until the required spatial dimensions are obtained (128, 128). 200 kernels of size (4 x 4) are used in each layer. The last three convolutional layers have a stride of 1 and have 200, 8 and 3 convolutional kernels respectively, to downsize the number of channels to the final output dimensions of (128, 128, 3) - corresponding to the image triplet Stokes I, Q and U. The CNN uses the training data (probe images from radiative transfer code) to learn the mapping from observables to images with an image based mean squared error (MSE) as the loss function. 

 \vspace{3mm} 

The network is trained on 10,000 training examples, using an automatic learning rate scheduler that commences at $10^{-4}$, and training is conducted until there is negligible improvement to the validation loss. A batch size of 32 was used. Dropout was used as regularization and tuned to 0.3. Leaky ReLU activation functions were used, the negative slope coefficient $\alpha$ was tuned to 0.25. Training of the CNN took approximately 4 hours on an NVIDIA GeForce RTX 4090.

\vspace{3mm}  

Stage 2 of PIRATES, an iterative fitting mechanism, is only applied at inference time and to a single observation at a time. First, stage 1 (the pre-trained CNN) is used to reconstruct polarized images from observables. The kernel weights of a copy of the pre-trained CNN are then further updated - wherein the objective function is now replaced with the MSE between the true interferometric observables and those recalculated from the pre-trained CNN images. Then commences stage 2 (iterative fitting) - the pre-trained CNN images are optimized with respect to their fit to the true interferometric observables. Since convolutional kernel weights -- rather than the image pixels -- are adjusted during iterative fitting, the reconstructions take place within the physics-informed feature space learned by the pre-trained CNN. 

\vspace{3mm}

At the commencement of stage 2, each copy of the CNN becomes a `single use' item - the pre-trained CNN must be freshly reloaded every time iterative fitting is deployed. Whilst iterative fitting continues to update the CNN weights, we refer to it as `fitting' and not `training', as iterative fitting does not continue to learn a generalizable mapping. Instead, the pre-trained convolutional kernel weights are used as initialization for the iterative fitting, which then optimizes a single image with respect to its observables. When running the iterative fitting, we used an adaptive learning rate scheduler commencing from a learning rate of 1e-5, with a conservative reduction rate of 0.8 and a patience of 5. To ameliorate computational overheads incurred when ending each training epoch, we populated each epoch with $\sim$75 instances of the input data, balanced against the requirement to regularly complete epochs to trigger the learning-rate scheduler (a limit which could be removed with custom-written code). Iterative fitting typically takes 5 - 10 minutes and 1500 - 2000 epochs to converge.
 
 \vspace{3mm}
 
 \subsection{Training and Validation `Probe' Datasets and Simulated Astrophysical Data}
  
 \vspace{3mm}  
 
Our CNN is pre-trained on a custom built dataset of polarimetric `probe' images and resulting polarimetric interferometric observables generated using radiative transfer code MCFOST \cite{Pinte2006, Pinte2009} (Figure \ref{vid:blobs_v2}). Choice of training data is crucial for imposing a physically meaningful prior of realistic spatial and polarization distributions. Our training data contains a distribution of scattering representations which feature minimal noise or unwanted artifacts from the radiative-transfer calculations, and have no bias towards any particular astrophysical structure. To achieve this balance, we perform radiative-transfer calculations for an ensemble of randomly generated dust-density volumes, each of which is made up of sets of three-dimensional-Gaussian probe-volumes of randomly chosen location, size and density, and situated around a randomly sized star. Each of these then produces corresponding 
polarized images and observables (Figure \ref{vid:blobs_v2}). Gaussian volumes spanning a range of sizes probe all spatial scales while avoiding aliasing due to their smoothness, discouraging pixel-level noise or sharp structures in image reconstructions. We convolve all images with a Gaussian kernel of full width half maximum of 1.5 pixels to help suppress pixel-scale stochastic noise arising from the discrete number of photons used within radiative transfer calculations and the edge artifacts from a discrete number of volumetric model cells.

\begin{video}[h!]
\begin{center} 
{\includegraphics[trim = 4.8cm 5.91cm 7.8cm 6.45cm, clip,  width= \textwidth]{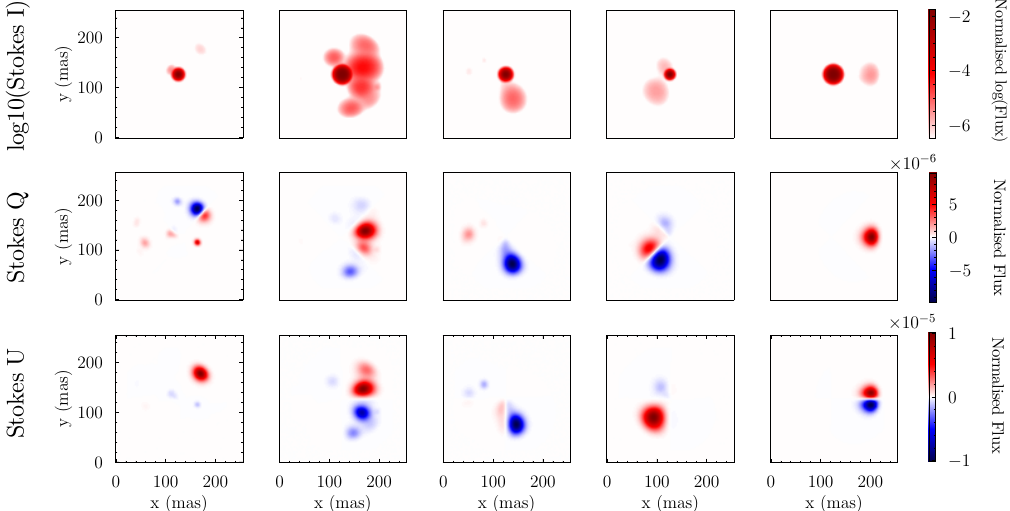}}
\end{center}
\caption{\label{vid:blobs_v2} Five examples of CNN pre-training probe image triplets. Each training example (column) consists of a (log10) Stokes I (first row), Q (second row) and U (third row) image. Images are calculated via a radiative transfer code, using an ensemble of randomly drawn Gaussian probe densities. The CNN learns a mapping between interferometric observables (visibilities and closure phases) and polarized image triplets, that is agnostic to specific astrophysical structure. Moreover, it learns to represent this mapping as a maximally parsimonious, physics-informed feature space.}
\end{video}
 Input and output data were both normalized prior to training. Inputs of interferometric observables are normalized with respect to their subclasses. Outputs of polarized images are also normalized with respect to their subclasses but require two intermediate modifications. Since the star is orders of magnitude brighter than the surface brightness of the dust, the logarithm of the Stokes I image is used whilst training the network, so that faint dust structures in Stokes I adequately influence loss calculations. In addition, all three images are normalized by the sum of flux in the (natural) Stokes I image, as is conventional in interferometric image reconstruction since the visibilities are independent of absolute flux.

\vspace{3mm}

Radiative transfer code was used to make 5500 pairs of polarized images and interferometric observables, which were then subject to augmentation. Images were randomly rotated and flipped, with appropriate Mueller matrices applied to maintain the polarimetric coordinate system, and interferometric observables were resampled from augmented images. A total of 11,000 examples were generated, 10,000 of which were used for training data and 1000 of which were reserved as validation data, which is used to independently evaluate the prediction accuracy of the CNN at the end of each pre-training epoch. Separately, prototypical circumstellar objects like disk and envelopes were injected into radiative transfer code, to create a set of simulated astrophysical data. This dataset was used to evaluate the performance of the iterative fitting, and was not used during training of the CNN (stage 1). More broadly, the simulated astrophysically dataset was used to assess how well PIRATES generalizes to spatially extended and astrophysically meaningful structures.

\vspace{3mm}
 
\section{Results and Discussion Part I - Architectural Tuning} \label{sect:results_1_general} 

This section summarizes the tuning of PIRATES' architecture. For PIRATES' final results and performance, see Results and Discussion Part II (Section \ref{sec:performance}). PIRATES hyper-parameters were tuned on the validation data to maximize the accuracy of final iteratively fit images, using the MSE and SSIM (structural similarity index measure) to quantify pixel-wise and structural accuracy.

\subsection{Stage 1 - CNN Architecture}\label{sect:results_1_CNN} 

Three architectural hyper-parameters were tuned: the number of convolutional layers, the number of convolutional kernels per layer, and the size of the convolutional kernels. We evaluated the impact of the number of layers and kernels simultaneously - conducting a grid search over reasonable values of both parameters. We maintained a `rectangular' network shape - adding the same number of kernels to every layer, with the exception of the final two convolutional layers which downsize the kernel number to the output channel size of 3 (Figure \ref{vid:final_F}). When varying the number of convolutional layers we were constrained by the requirements of adequate receptive field and adequate up-sampling to the final output image size of (128, 128). Up-sampling was achieved by using early convolutional layers with a stride of 2, and layers beyond layer 5 have a stride of 1. 

\begin{video}[htbp]
\centering 
{\includegraphics[trim = 0cm 0cm 0cm 0cm, clip,  width= 0.99\textwidth]{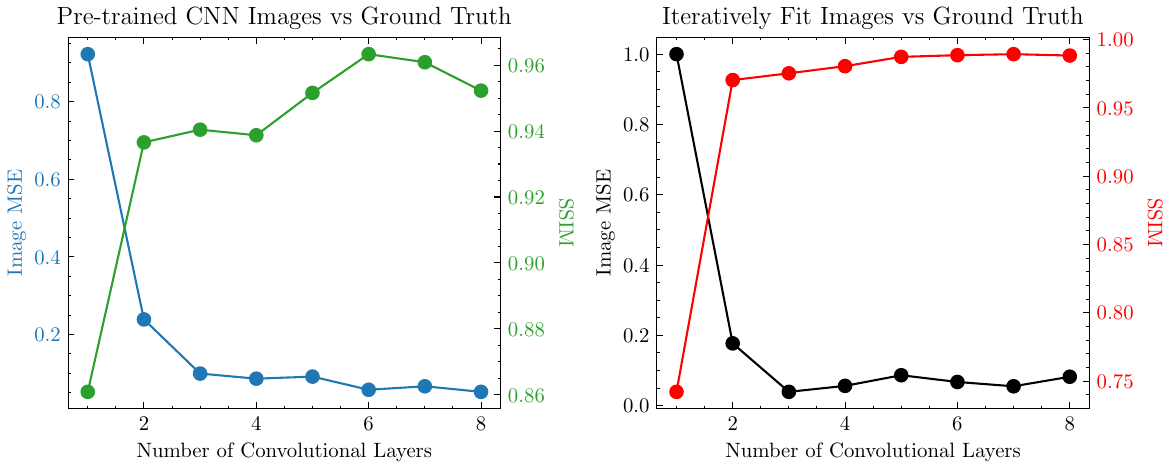}} 
\caption{\label{vid:grid} The influence of changing the number of intermediate convolutional layers on the performance of the a) the pre-trained CNN only (left panel), and b) the entire image reconstruction algorithm - CNN + iterative fitting (right panel), with respect to the validation `probe' data. The influence of increasing layers is evaluated using both the image MSE and SSIM. Adding intermediate convolutional layers improves performance with diminishing returns after $\sim$ 6 layers. In all examples, two final convolutional layers are used to downsize the channel size to 3.}
\end{video}

We found that increasing the number of convolutional kernels provided enhanced validation performance with diminishing returns after approximately 200 kernels were added to each layer, and adopted 200 kernels per layer for the remainder of this analysis. We restricted ourselves to even convolutional kernel sizes, to avoid checkerboard noise associated with the combination of odd kernel sizes and stride 2 convolutions \cite{Tanaka}, and found that (4 x 4) was the optimal kernel size. The number of convolutional layers had the strongest influence on the accuracy of reconstructed images - see Figure \ref{vid:grid}. The best final reconstruction accuracy after iterative fitting was found to be at $\sim$6 intermediate layers, with no improvements seen beyond that, suggesting that this number of intermediate layers is sufficient to represent the complexity of this feature space (Figure \ref{vid:grid}, right). 
 \vspace{3mm} 

This trend was also consistently observed with simulated astrophysical data. As convolutional layers are added, the pre-trained CNN images themselves converge to images with modest detail, but more critically, low amounts of noise (Figure \ref{vid:grid_ims}, top row).
As the pre-trained CNN initializes the iterative fitting, this suggests that initialization from a lower resolution but less noisy feature space is more advantageous for the iterative fitting. Using a larger number of layers results in final iteratively fit images that are more accurate and contain minimal noise (Figure \ref{vid:grid_ims}, bottom row, also see ground truth image on far right). We thus adopt six intermediate convolutional layers - five of stride 2, and one of stride 1, as an optimum architectural configuration for the CNN of PIRATES.

\begin{video}[htbp]
\centering 
{\includegraphics[trim = 0.1cm 6.5cm 1.20cm 3.8cm, clip,  width= 0.99\textwidth]{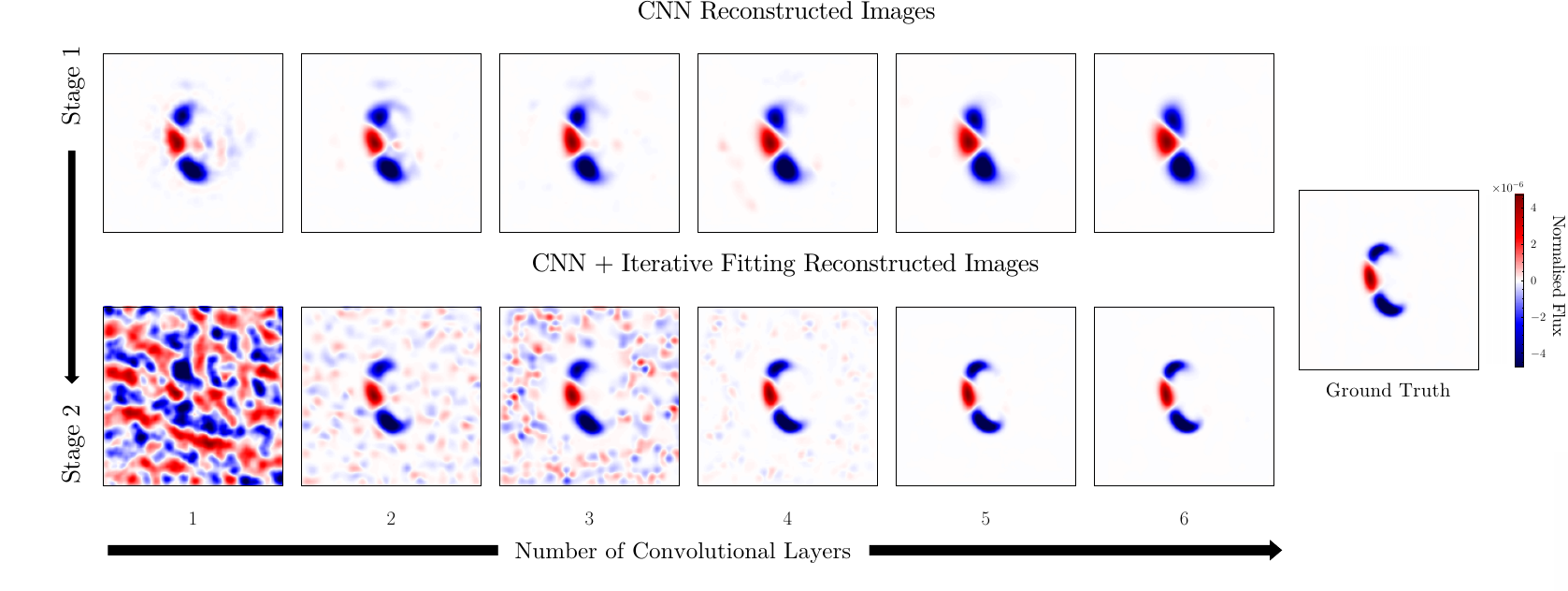}}
\caption{\label{vid:grid_ims} The influence of increasing the number of intermediate convolutional layers on the images reconstructed at each stage of PIRATES (top row - stage 1, bottom row - stage 2), demonstrated for a simulated astrophysical example of an inclined circumstellar disk. As convolutional layers are added, pre-trained CNN images converge to modest detail and low noise versions of the ground truth, which will subsequently produce the most accurate and low noise final iteratively fit images (compare to ground truth image in bottom right).}
\end{video}

 \subsection{Stage 2 - Iterative Fitting Architecture}\label{sect:results_1_it} 
 
 \vspace{3mm}  

The iterative fitting is tuned to minimize the residuals between the true observables and those recomputed from the pre-trained CNN images, whilst also ensuring that improvements to the pre-trained CNN images are self-consistent and do not introduce significant spurious noise or structure. The improvement which iterative fitting yields to the observables is used explicitly as the loss function (MSE of observables), however we also monitored the image MSE and the SSIM as metrics to quantify both pixel wise and structural improvements to the images. The iterative fitting process can be tuned by controlling how many CNN layers are iteratively fit - as opposed to having their weights `frozen' to pre-trained values for the duration of the iterative fitting. To assess this, we investigated the impact of freezing layers in both directions (shallow-deep). We found that iterative fitting has the best performance on validation data when all layers are iteratively fit and none are frozen - this optimum is measurable in the improvement to observables, as well as both image based MSE and SSIM scores (Figure \ref{vid:it_layers_num}).

\begin{video}[htbp]
\centering 
{\includegraphics[trim = 0cm 0cm 0cm 0cm, clip,  width= 1.0\textwidth]{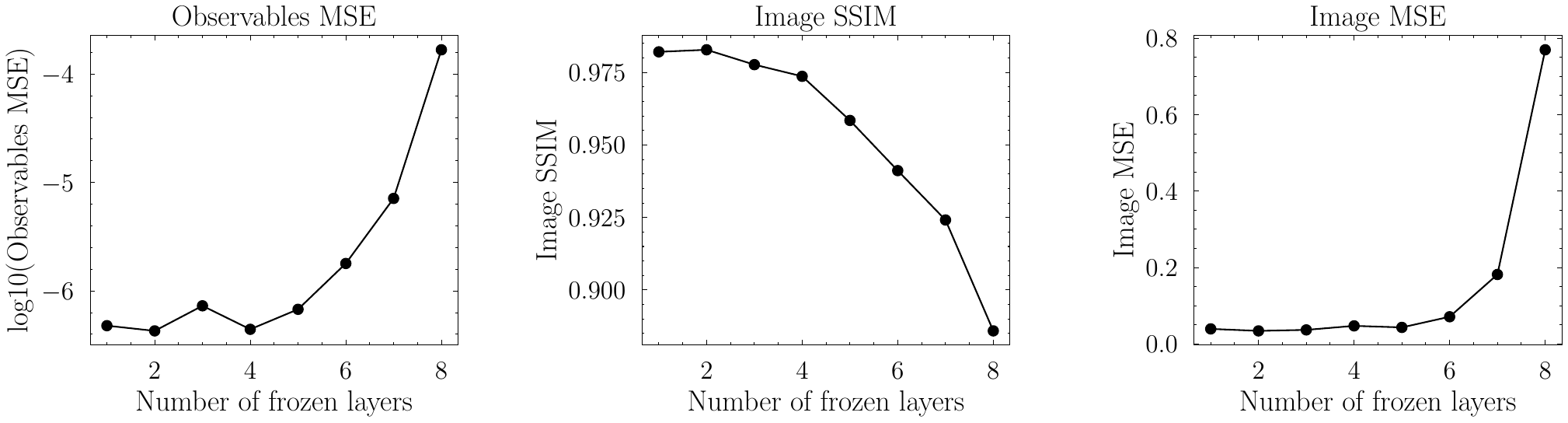}}
\caption{\label{vid:it_layers_num} The influence of freezing convolutional layers during iterative fitting. Here the results are plotted for freezing layers starting from the shallowest layers to the deepest layers, though the results are the same for deep-shallow. The iterative fitting performs optimally with the maximum number of degrees of freedom - where all layers are iteratively fit, and none are frozen to pre-trained values.}
\end{video}

  \vspace{3mm}

\section{Results and Discussion Part II - Performance of the two-stage framework on simulated astrophysical examples}\label{sec:performance}

\vspace{3mm}

Here we demonstrate the performance of PIRATES using a our simulated astrophysical dataset of disks, spirals and complex circumstellar environments, generated using the MCFOST radiative transfer code. These examples encode the astrophysical complexity we anticipate underlie VAMPIRES NRM (and other polarized-interferometric) datasets - optical depth, non-azimuthal scattering and significant geometric asymmetry. In making these examples, we drew inspiration from real observations of the circumstellar environments of evolved stars like Betelgeuse \cite{Cannon2023} and $\mu$ Cephei \cite{Borgne1987}, debris disks like HR 4796 \cite{Lagrange2012, DeRegt2024a}, and spiral armed disks like HD 135344B\cite{DeRegt2024a}, HD 142527 \cite{DeRegt2024a} and MWC 758 \cite{DeRegt2024a}.  Figures \ref{vid:eg_1} - \ref{vid:eg_5} demonstrate the image reconstruction performance of PIRATES on a sub-set of the simulated astrophysical dataset, and include the predictions from the pre-trained CNN alone (the `initial guess') (rows 1 and 2) and the final predictions after iterative fitting (rows 3 and 4) - for both the images and interferometric observables.

\begin{video}[htbp]
\centering 
{\includegraphics[trim = 9.1cm 2.8cm 11.1cm 3.0cm, clip,  width=0.82\textwidth]{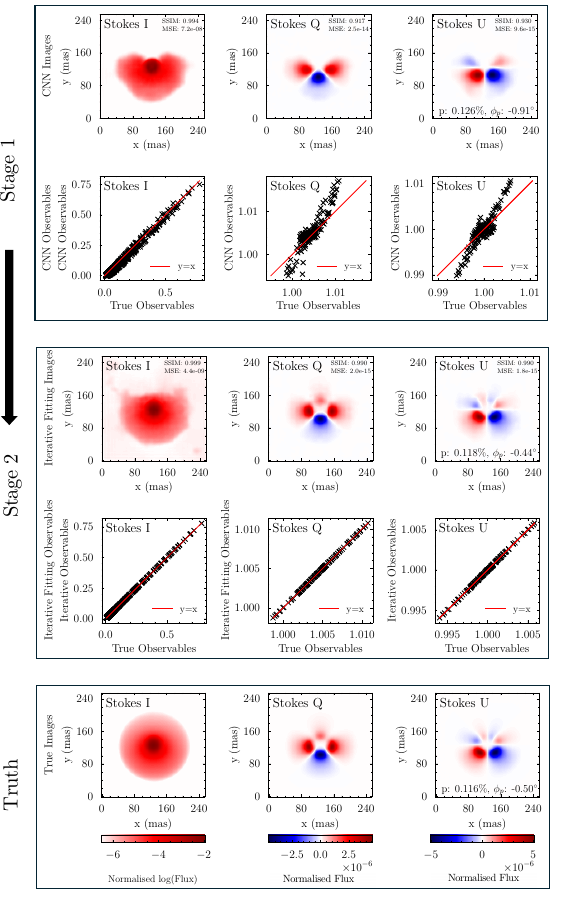} }  
\caption{\label{vid:eg_1}  PIRATES performance on an inclined model circumstellar disk, with a non-azimuthal scattering feature on the top edge of the disk.}
\end{video}

\begin{video}[htbp]
 \centering 
{\includegraphics[trim = 9.8cm 2.6cm 9.5cm 2.5cm, clip,  width=0.85\textwidth]{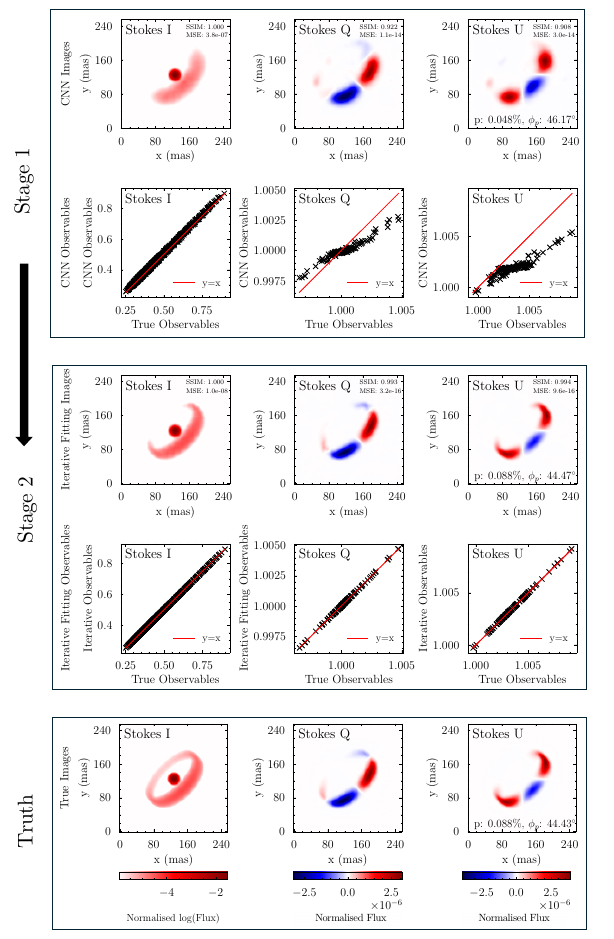} }  
\caption{\label{vid:eg_7} PIRATES performance on an inclined model circumstellar disk. The effects of large Mie grains are evident - the back of the disk is shadowed and the image is dominated by strong forward scattering within the disk's front.}
\end{video}

\begin{video}[htbp]
 \centering 
{\includegraphics[trim = 9.0cm 2.5cm 10.5cm 2.5cm, clip,  width=0.85\textwidth]{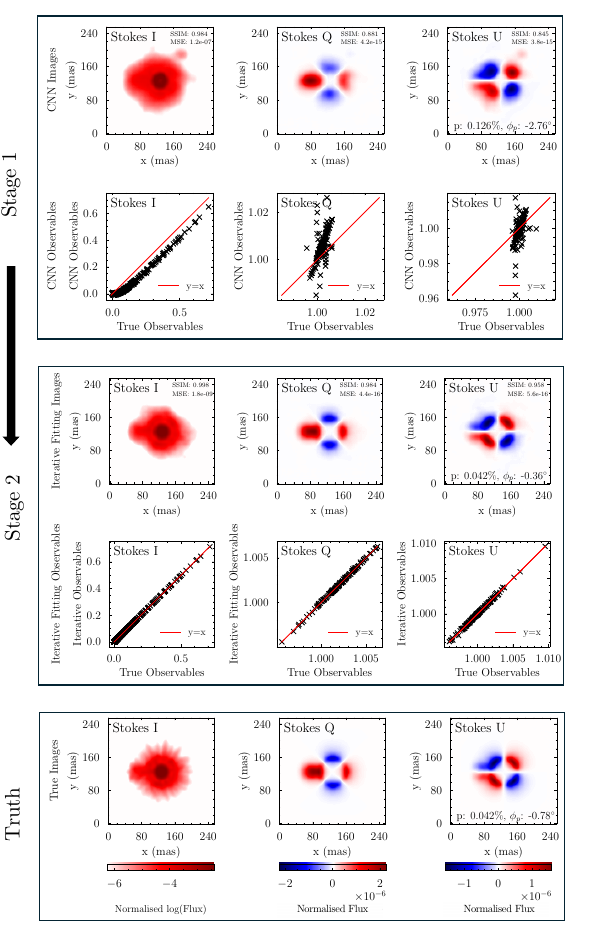} }  
\caption{\label{vid:eg_4}  PIRATES performance on a model clumpy, inhomogeneous circumstellar environment typical of evolved stars. }
\end{video}

\begin{video}[htbp]
 \centering 
{\includegraphics[trim = 9.0cm 2.5cm 11.0cm 2.5cm, clip,  width=0.82\textwidth]{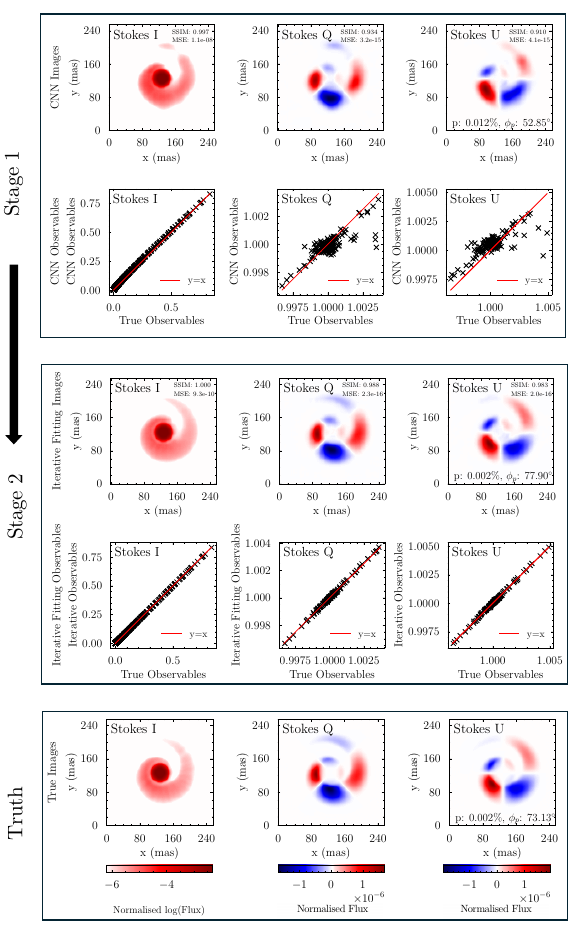} }  
\caption{\label{vid:eg_9}   PIRATES performance on a model dust spiral. }
\end{video}

\begin{video}[htbp]
 \centering 
{\includegraphics[trim =  9.8cm 2.5cm 9.5cm 2.5cm, clip,  width=0.85\textwidth]{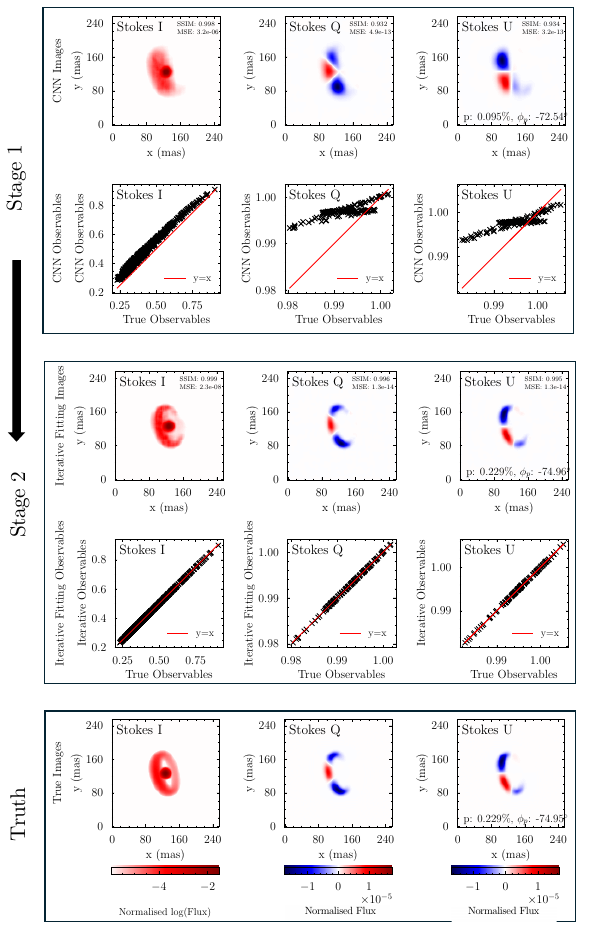} }  
\caption{\label{vid:eg_6} PIRATES performance on a model circumstellar disk. }
\end{video}

\begin{video}[htbp]
 \centering 
{\includegraphics[trim =9.8cm 2.5cm 9.5cm 2.5cm, clip,  width=0.85\textwidth]{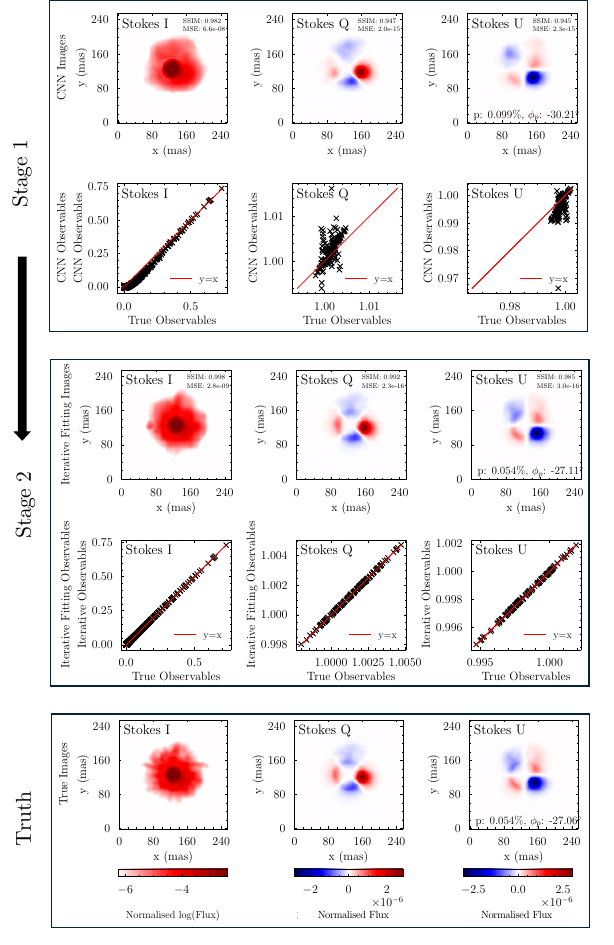} }  
\caption{\label{vid:eg_5} PIRATES performance on a model dusty circumstellar environment typical of evolved stars.  }
\end{video} 

\vspace{3mm}

It can be seen that the predictions from the pre-trained CNN alone give surprisingly accurate results, with broad features of the true image being reproduced (Figures \ref{vid:eg_1} - \ref{vid:eg_5},  rows 1). Finer detail is missing, and the appearance of overly connected structure or excess material in the image centers occurs. The observables resampled from the CNN images are relatively poor fits to those sampled from the ground truth images (Figures \ref{vid:eg_1} - \ref{vid:eg_5},  rows 2). As iterative fitting proceeds, observables are typically optimized to well below realistic observational levels of uncertainty ($O(1\%)$) (Figures \ref{vid:eg_1} - \ref{vid:eg_5},  rows 4); simultaneously the reconstructed images are improved on both a pixel wise and structural level (Figures \ref{vid:eg_1} - \ref{vid:eg_5},  rows 3), indicated by improved MSE and SSIM scores. As is visually apparent in Figures \ref{vid:eg_1} - \ref{vid:eg_5}, all astrophysical examples see clear improvement in both image accuracy and observables accuracy as a result of the iterative fitting. The final reconstructed images are high fidelity and noise free, accurately resembling the ground truths.

  \vspace{3mm}

There are two consistent styles of refinement which iterative fitting makes to the reconstructed images. Firstly, iterative fitting enhances the resolution of the CNN pre-trained images to match the information contained in the interferometric observables. This effect is evident across all simulated astrophysical examples - disk profiles are narrowed (Figures \ref{vid:eg_7}, \ref{vid:eg_1}, \ref{vid:eg_6}), the inner cavities of clumpy circumstellar dust shells are widened (Figure \ref{vid:eg_4}, \ref{vid:eg_5}) to reveal inner dust radii. The second effect which iterative fitting has on the reconstructed images is that it is capable of reconstructing material that was originally missing from the pre-trained CNN reconstructions, including - fine non-azimuthal lobes of circumstellar disks (Figure \ref{vid:eg_1}) and the full length of a spiral tail (Figure \ref{vid:eg_9}). Iterative fitting is also capable of removing structure that was incorrectly present within the pre-trained CNN images; a faint circle of dust in the center right of Figure \ref{vid:eg_4}, some structure in the center left of Figure \ref{vid:eg_7}. These image improvements are quantified by improved MSE and SSIM scores for each image.

\vspace{3mm}

Refinements which the iterative fitting makes to images are mutual and consistent across all three Stokes images - both locally and globally, demonstrated not only by the mutual improvement in polarized structure within the images, but also by measuring the total polarization $p= \frac{\sqrt{Q^2 + U^2}}{I}$ and polarization angle $p_\phi =\frac{1}{2} \arctan{\frac{U}{Q}}$, as annotated in the right column of each example (Figures \ref{vid:eg_1} - \ref{vid:eg_5}). Improvements to the images occur without the introduction of spurious noise or structure, as the pre-trained CNN is optimized as a modest detail but low noise initialization and feature-space for the iterative fitting.

\section{Performance with random noise}

We also demonstrate that PIRATES is compatible with the presence of realistic amounts of random error. To investigate the impact of noise, we inject the interferometric observables of our simulated astrophysical examples with error that matches the signal to noise ratio previously seen in on sky NRM datasets (Figure \ref{vid:eg_14} columns 1 and 2). With no adaptation of the PIRATES network, we find that stage 1 (the CNN) still does a good job at providing moderate resolution and low noise initializations for the iterative fitting (Figure \ref{vid:eg_14}, row 1, columns 3-5), as it does for noise-free examples (Figures \ref{vid:eg_1} - \ref{vid:eg_5}). However, when stage 2 is deployed and images are iteratively fitted with respect to their (now noisy) observables, significant amounts of noise are now introduced into the images, as may be anticipated (Figure \ref{vid:eg_14}, row 2, columns 3-5). Thus, to perform optimally on noisy real datasets, PIRATES requires adaptation to reduce the noise and spurious structure that iterative fitting may inject into the reconstructed images as a consequence of noise.

\begin{video}[htbp]
 \centering 
{\includegraphics[trim = 2.30cm 3.0cm 1.99cm 2.85cm, clip,  width=\textwidth]{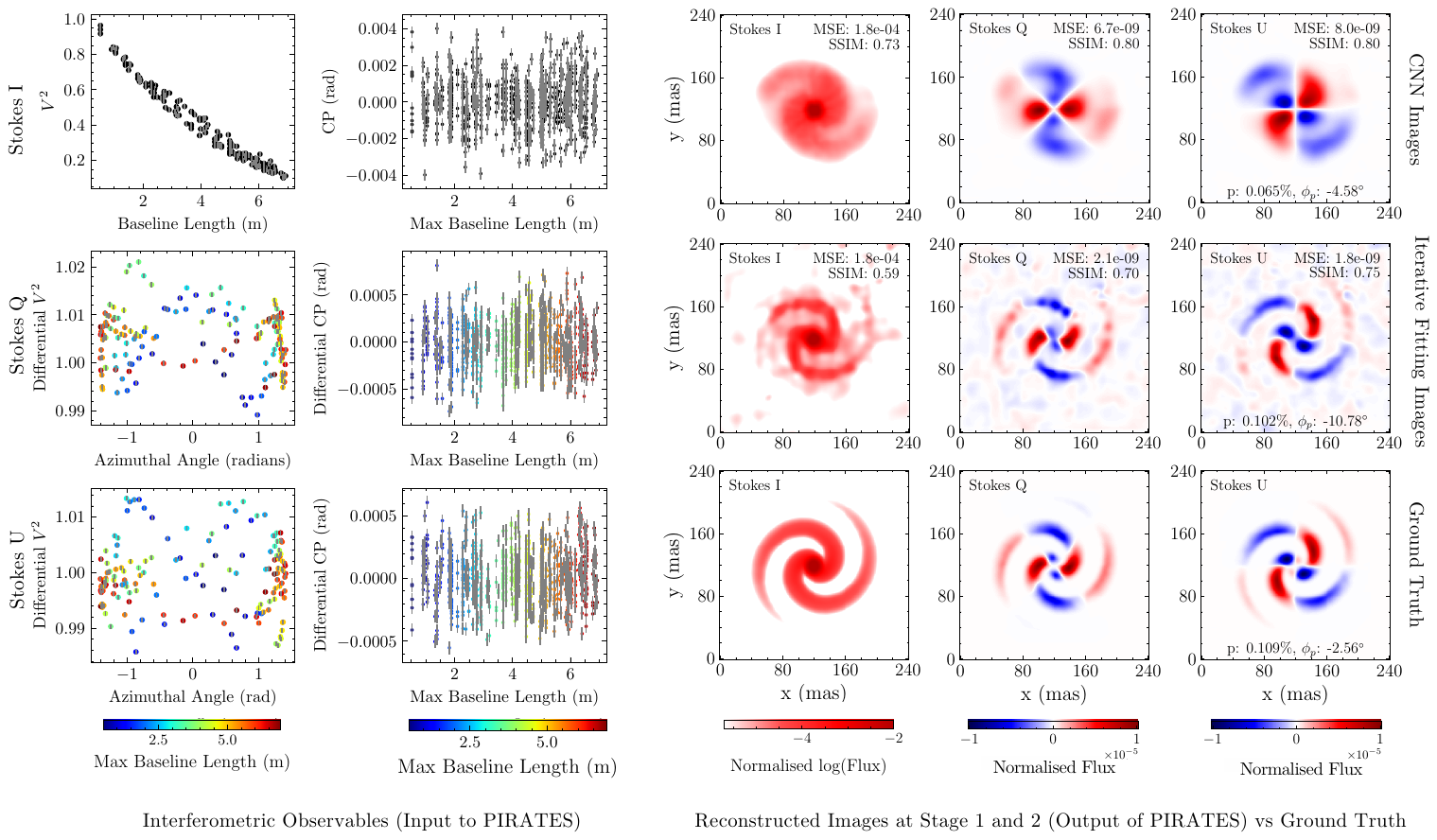}}
\caption{\label{vid:eg_14} The performance of PIRATES with signal to noise consistent with recent VAMPIRES NRM data, with no algorithmic treatment to constrain the influence of noise. Visibilities and closure phases with injected noise and corresponding error bars are plotted in columns 1 and 2. Stage 1 (CNN) does a good job of a moderate resolution and low noise reconstruction of the ground truth (top row, columns 3-5), however, significant amounts of noise are introduced into the images during the iterative fitting (middle row, columns 3-5). The ground truth images are displayed in the bottom row, columns 3-5.}
\end{video}

  \vspace{3mm}

In the presence of noise, the loss function of the iterative fitting is updated to a chi-squared $\chi^2$ statistic, allowing the iterative fitting to more heavily weight the reconstruction of features represented by higher signal to noise observables. In the spirit of the `implicit regularization' functionality of PIRATES, rather than regularizing the reconstructed images in pixel space like traditional image reconstruction codes, we regularize the CNN's latent feature space - restricting its functional complexity and encouraging lower complexity solutions which minimize noise. There are many ways to implement latent feature space regularization - here we explore one promising solution, but do not assert this is the only solution.
 
  \vspace{3 mm}

To provide the additional regularization required to prevent stage 2 (iterative fitting) from introducing noise to reconstructed images, we use an Elastic Weights Consolidation penalty during the iterative fitting stage. Elastic Weights Consolidation (EWC) is useful where retaining historical information benefits future predictive accuracy \cite{EWC}. It is employed by introducing a continuous penalty for the gradual `forgetting' of network weights which prior information indicates will improve the accuracy of a sequential prediction. In the context of PIRATES, we introduce a penalty during the iterative fitting if iteratively fit weights are modified too much from their pre-trained CNN values. The choice of EWC as regularization is guided by intuition that the CNN's predictions are already reasonably good, and more crucially, low noise representations of the ground truth. With an EWC penalty, changes to the pre-trained CNN weights are only made where they strongly improve the iteratively fit image's fit to observables, discouraging non-critical changes to weights that may introduce spurious noise or structure. The elegance of EWC is that when it is applied to an extreme level, it simply converges to the pre-trained CNN prediction. This gives a clear indication where over-regularization has occurred, making it robust to tune.

\vspace{3mm}

To create a penalty using EWC, we are first required to quantify the `importance' of each pre-trained CNN parameter, $\theta_j$. We then use this `importance' to weight the contributions of each network parameter to the EWC penalty. To quantify the importance of each network parameter, we derive a heuristic for the diagonal of the empirical Fisher Information matrix as is standard practice when defining an EWC penalty for machine learning \cite{EWC}. 

\vspace{3mm}

To do so, we first compute an average loss $\mathcal{L}_b$ over a batch of N input-output pairs ($x_i, y_i$) (Equation \ref{eq:L}), where $\ell$ is the pre-trained CNN loss function and $f_\theta$ denotes the CNN, constructed of network parameters $\theta_j$ (weights, biases).

\begin{equation}
   \mathcal{L}_b = \frac{1}{N} \sum_{i=1}^N \ell(y_i, f_\theta(x_i))
\label{eq:L}
\end{equation}

We then use backpropagation to compute the gradient $g_j^{(b)}$ of $\mathcal{L}_b$ with respect to each model parameter $\theta_j$ (Equation \ref{eq:g}).

\begin{equation}
    g_j^{(b)} = \frac{\partial \mathcal{L}_b}{\partial \theta_j}
\label{eq:g}
\end{equation}

\vspace{3mm}

We use the average squared gradients over $M$ batches of training data, $Fj$, as a heuristic for the Fisher Information of parameter $j$, (Equation \ref{eq:Fish}):

\begin{equation}
    F_j =  \frac{1}{M} \sum_{b=1}^{M} \left( g_j^{(b)} \right)^2 = \frac{1}{M} \sum_{b=1}^{M} \left( \frac{\partial \mathcal{L}_b}{\partial \theta_j} \right)^2
\label{eq:Fish}
\end{equation}

 $F_j$ is used to weight the squared difference between current $\theta_j$ (iterative fitting) and previous $\theta_j^*$ (pre-trained CNN) network parameters (Equation \ref{eq:ewc_loss}), penalizing changes to weights which were more important to the CNN's original reconstructions. 

\begin{equation}
\label{eq:ewc_loss}
EWC =  \lambda \sum_{i} F_j \left( \theta_j - \theta_j^{*} \right)^2
\end{equation}

The strength of the EWC penalty is tuned via hyperparameter $\lambda$, the final loss function of the iterative fitting, $L$, is given by Equation \ref{eq:it_loss} \cite{EWC}.

\begin{equation}
\label{eq:it_loss}
L = \chi^2 +  \lambda \sum_{j} F_j \left( \theta_j - \theta_j^{*} \right)^2
\end{equation}

\begin{video}[htbp]
 \centering 
{\includegraphics[trim = 8.05cm 7.2cm 12.25cm 3.45cm, clip,  width=\textwidth]{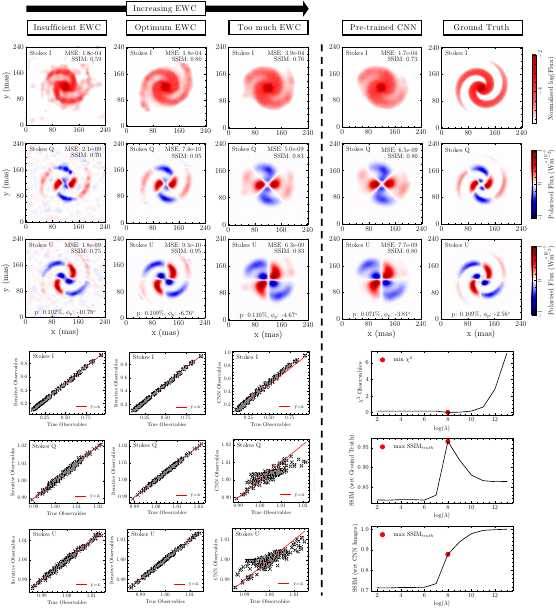}} 
\caption{\label{vid:eg_10sc} The impact of increasing EWC strength via $\lambda$ (Equation \ref{eq:it_loss}). With low $\lambda$ (col 1), significant noise is present in reconstructed images. As $\lambda$ is increased (col 2), the noise present in images is substantially reduced and the SSIM of the reconstructed images is improved relative to the ground truth (col 5). When $\lambda$ is applied to an extreme level (col 3), reconstructions do not deviate from the CNN pre-trained images (col 4). $\lambda$ may be tuned by monitoring: the SSIM between iteratively fit images and the pre-trained CNN images (col 4-5, row 6), and the fit to interferometric observables (col 4-5 row 4). We found that the optimal value of $\lambda$ (where the SSIM between the iteratively fit images and the ground truth is maximized) (col 4-5, row 6) was correlated with an optimized fit to observables (col 4-5, row 4).}
\end{video}

The effect of increasing the strength of EWC regularization via $\lambda$ is visually striking. The results of iterative fitting with insufficient $\lambda$ are noisy images (Figure \ref{vid:eg_10sc}, col 1). When $\lambda$ is of optimal strength, the iterative fitting converges to images with minimal noise (Figure \ref{vid:eg_10sc}, col 2), and the fit to interferometric observables is simultaneously optimized (Figure \ref{vid:eg_10sc}, col 5, row 4). The SSIM scores of iteratively fit images are improved, as are the reproduction of polarization statistics (inset in Stokes U panels). As expected, when EWC regularization is applied to an extreme level, the iteratively fit images do not vary much from the pre-trained CNN images (Figure \ref{vid:eg_10sc}, col 3), nor is there improvement in the fit to observables from the pre-trained CNN initialization.

\vspace{3mm}

The influence of the EWC penalty on image reconstruction is easiest to assess by monitoring the SSIM between the pre-trained CNN and iteratively fit images, as a function of $\lambda$ (Figure \ref{vid:eg_10sc}, col 4-5, row 6). Small values of $\lambda$ result in noisy iterative fitting images, which are most structurally dis-similar to the CNN images. As $\lambda$ is increased, the SSIM score increases as the algorithm is encouraged to fit the observables to real structure and not to noise, now maintaining network parameters which were highly informative in the pre-trained CNN. However, eventually $\lambda$ becomes so large that the iteratively fit images barely deviate from the CNN images (and hence the SSIM converges to $\sim$ 1).

\vspace{3mm}

To locate an optimum value of $\lambda$, we found that a useful heuristic was to monitor the fit to interferometric observables. For our simulated astrophysical examples, with known ground truths, we know a-priori the optimal value of $\lambda$, as we can measure the SSIM between the iteratively fit images and the ground truth (Figure \ref{vid:eg_10sc}, col 4-5, row 5). We found that the value of $\lambda$ where the goodness of fit of interferometric observables was optimized was correlated with the value of $\lambda$ that maximized the SSIM of the iteratively fit images and their ground truths (Figure \ref{vid:eg_10sc}, col 4-5, row 4 and 5). Initially, increasing $\lambda$ improves the fit to observables (Figure \ref{vid:eg_10sc}, col 4-5, row 4). As $\lambda$ is increased, the sub-space of possible image solutions shrinks - as noisy solutions are excluded by regularization. As such, we interpret the improvement which increasing $\lambda$ yields to the observables as a consequence of more stable and uniquely defined convergence, towards solutions that are free of noise that may perturb the fit to observables. In summary, in real applications with no known ground truth, increasing $\lambda$ until the goodness-of-fit of the interferometric observables begins to exceed the levels determined by observational error is recommended as a useful heuristic, along with simple inspection of the images \cite{Hansen1992, Vandenberghe}.

\vspace{3mm}

\section{Conclusions}\label{sect:conclusions} 

We have presented PIRATES, the first image reconstruction algorithm for optical polarimetric interferometry, which uniquely performs image reconstruction in a learned, physics-informed feature space rather than in pixel space. PIRATES' dual structure - a pre-trained CNN and an iterative fitting mechanism - allow the algorithm to parsimoniously reconstruct high fidelity polarized images which accurately reproduce polarized, interferometric observables. We have demonstrated that PIRATES accurately reconstructs polarized images of a diverse range of astrophysically realistic radiative transfer models, including disks, spirals and clumpy circumstellar environments. 

\vspace{3mm}

We found that optimization of PIRATES dual stage structure required a CNN which constructs modest detail but low noise versions of the ground truth, providing a physics-informed feature space to be used as initialization for iterative optimization in the second stage. The iterative fitting stage is capable of improving the reconstructed images consistently across all polarimetric states (Stokes I, Q and U), without introducing noise or spurious structures, simultaneously improving the fit to observables to within realistic observational uncertainty. PIRATES implicitly regularizes reconstructed images via a physically meaningful prior that the CNN learns from pre-training data (produced via radiative transfer models), and the structurally derived inductive bias that forces image updates made by iterative fitting to be statistically consistent across the image domain. In the presence of realistic amounts of noise in interferometric observables, we found that introducing Elastic Weights Consolidation as a latent feature space regularizer was able to prevent the introduction of noise into images during the iterative fitting, and that a single parameter to adjust the strength of regularization was sufficient and robust in tuning the algorithm.

\vspace{3mm}

In future work we intend to upgrade PIRATES to streamline flexible polarimetric coordinates and varying (u,v) coverage, by adding an auto-encoder to the beginning of our existing algorithm. Additionally, our treatment of noise only includes random error - strategies for treating systematic sources of error that are common to polarized optical systems are in development. The performance of our algorithm demonstrates the power of dual-structure machine learning algorithms, and suggests they may more broadly be a useful technique for contexts with complex regularization constraints, beyond just those of polarized interferometry.

\vspace{3mm}

 \section{Data and Code availability}

 Code for PIRATES is available on GitHub: \href{https://github.com/LucindaLilley/PIRATES.git}{github.com/LucindaLilley/PIRATES}.

 \section{Acknowledgments}

 \vspace{3mm} 

 The authors wish to recognize and acknowledge the very significant cultural role and reverence that the summit of Maunakea has always had within the indigenous Hawaiian community, and are most fortunate to have the opportunity to conduct observations from this mountain.

\vspace{3mm}

Based on data collected at Subaru Telescope, which is operated by the National Astronomical Observatory of Japan. The development of SCExAO is supported by the Japan Society for the Promotion of Science (Grant-in-Aid for Research \#23340051, \#26220704, \#23103002, \#19H00703, \#19H00695 and \#21H04998), the Subaru Telescope, the National Astronomical Observatory of Japan, the Astrobiology Center of the National Institutes of Natural Sciences, Japan, the Mt Cuba Foundation and the Heising-Simons Foundation. 

\vspace{3mm}

Majority funding for the development of the VAMPIRES instrument came from the Australian Research Council (FT100100953; DP140104065).

\vspace{3mm}

LL wishes to acknowledge Adam Taras, Benjamin Pope, Louis Desdoigts and Max Charles for their helpful discussions and advice over figures.

 \vspace{3mm}

An earlier version of this work was submitted as a Conference Proceedings at SPIE Astronomical Telescopes + Instrumentation, 2024, in Yokohama, Japan \cite{Lilley2024}.

\section{Disclosure Statement}

The authors declare there are no financial interests, commercial affiliations, or other potential conflicts of interest that have influenced the objectivity of this research or the writing of this paper.


\bibliographystyle{plain}  

\end{document}